# Sur le principe conceptuel fondamental de la mécanique quantique

## Frédérique Laurent & François-Igor Pris

**Résumé**

Le "principe conceptuel fondamental" de la mécanique quantique proposé par A. Zeilinger, selon lequel un système élémentaire est porteur d'un bit d'information, est un principe idéaliste qui doit être remplacé par un principe réaliste de contextualité. Les propriétés spécifiques des systèmes quantiques sont une conséquence de l'impossibilité de parler d'eux sans référence à des moyens de leur identification et, par conséquent, à un contexte dans lequel ces moyens sont employés. En particulier, pour expliquer la corrélation quantique, il n'est pas nécessaire de supposer sa non-localité. Les événements quantiques corrélés sont liés les uns aux autres de manière causale. Il ne s'agit pas d'une causalité classique, mais quantique, exprimée par une fonction d'onde intriquée. Dans la mesure, telle ou telle corrélation particulière n'apparaît pas; dans la mesure, elle est identifiée dans le contexte. Contrairement au principe informationnel de Zeilinger, le principe de contextualité l'explique de manière réaliste.

**Mots clés**: mécanique quantique, information, idéalisme, réalisme, contexte, problème de suivre une règle, corrélation quantique.

## 1. Introduction

En physique contemporaine, et en particulier dans le cadre de certaines interprétations informationnelles de la mécanique quantique, l'idée que l'information est une caractéristique fondamentale de l'univers s'est imposée. On dit aussi que l'univers (la nature) est un immense ordinateur ou programme. De même, sur les pas de Galilée, les scientifiques modernes pensaient que le livre de la nature était écrit dans le langage des mathématiques. Selon Husserl, «avec Galilée commence (…) la substitution de la nature idéalisée à la nature sensible préscientifique» [1, p. 247]. Il reproche à Galilée de confondre la réalité avec la méthode mathématique de sa représentation: «C'est le vêtement d'idées qui nous fait prendre pour l'être véritable ce qui est une méthode» [1, p. 248]. Du point de vue d'une science pratique, ce type de projection de modèles sur la réalité est acceptable, puisque la norme fondamentale de la science, comme la physique, est la vérité. Mais du point de vue philosophique, une confusion de l'idéal (conceptuel) – le modèle est idéal – et du réel (les choses elles-mêmes) est inacceptable. C'est l'une des raisons pour lesquelles des débats philosophiques ont lieu. Les discussions sur l'interprétation de la mécanique quantique, comme on le sait, se poursuivent depuis près de cent ans. Nous proposons d'examiner l'interprétation informationnelle de la mécanique quantique dans la perspective du réalisme contextuel de provenance wittgensteinien, qui établit une distinction entre les catégories du réel et de l'idéal [2–4]. Nous analysons le «principe informationnel» de Zeilinger, qu'il a proposé comme un «principe conceptuel fondamental» pour la mécanique quantique [2]. Nous affirmons qu'il s'agit d'un principe idéaliste, qui devrait être remplacé par un principe réaliste de contextualité.

## 2. Le réalisme contextuel

Dans le cadre d'un réalisme contextuel, les concepts, les règles, les normes, les informations, les connaissances appartiennent à la catégorie de l'idéal [5–9]. Si les concepts étaient réels, ils ne pourraient pas saisir la réalité, ou il faudrait quelque chose d'«irréel» pour les saisir eux-mêmes comme une réalité particulière. Ils ne sont réels que dans le sens où ils sont ancrés dans la réalité, ont des conditions réelles d'existence et d'efficacité (d'application). L'idéal est un «mouvement» dans la réalité, qui transcende les limites du sensible. Dans la cognition, nous appliquons des concepts, portons des jugements, identifions des éléments de la réalité, mais nous ne les construisons ni ne les transformons. L'idéal ne peut pas affecter le connaissable. Ainsi, nous devons rejeter les interprétations subjectivistes de la mécanique quantique, selon lesquelles l'observateur influence le système quantique observé. Il est également faux de parler d'une corrélation indépassable entre l'observateur quantique et le système quantique observé, comme si ce n'était pas la nature quantique elle-même, mais seulement les choses-quantiques-pour-nous qui nous étaient accessibles.

Selon notre réalisme contextuel wittgensteinien, toute théorie scientifique bien établie et confirmée, y compris la mécanique quantique, a un domaine d'applicabilité. Dans ce domaine, elle joue le rôle d'une règle (norme) wittgensteinienne pour mesurer la réalité, elle a une validité logique. Par conséquent, elle n'est pas falsifiable. La structure du problème de la mesure quantique est celle du problème wittgensteinien de suivre une règle. L'écart entre le formalisme quantique et la réalité est comblé dans le jeu de langage de son application, c'est-à-dire dans la pratique, dans le contexte. Dans ce cas, comme l'écrit J. Benoist, «context is not so much an external constraint on meaning—as if reality, so to speak, struck meaning from the outside—as it is the manifestation of meaning being *effectively rooted* in reality, as well as something that contributes to the constitution of meaning itself. If meaning does not have to 'make contact' with reality, that is because it is already active as a genuine normative 'move' within the space of reality» (le contexte n'est pas tant une contrainte externe sur le sens – comme si la réalité, pour ainsi dire, devait frapper le sens de l'extérieur - qu'une manifestation de ce que le sens est effectivement enraciné dans la réalité, et de ce qui contribue à la constitution du sens lui-même. Si le sens n'a pas besoin d'entrer en 'contact' avec la réalité, c'est parce qu'il est déjà actif comme un véritable 'mouvement' normatif dans l'espace de la réalité) [9, p. xii–xiii]. La soi-disant réduction de la fonction d'onde n'est pas un processus physique. Elle est due à un changement de contexte. En un sens, notre point de vue est un retour à l'interprétation de Copenhague de la mécanique quantique, ajustée dans le cadre du réalisme contextuel [10-12]. Nous l'appelons «réalisme contextuel quantique».

### 3. Le réalisme contextuel et le QBisme

Notre point de vue que la théorie quantique est la règle (norme) Wittgensteinienne présente des similitudes avec le QBisme. Selon le QBisme – le bayésianisme subjectif (personnaliste) quantique – la règle de Born et la théorie quantique sont normatives (prescriptives) et non descriptives; elles ne sont pas des descriptions de la réalité objective, existant indépendamment du sujet et de l'usage du langage théorique [13-15]. Le QBisme interprète pourtant la mesure quantique comme une interaction du sujet avec le système quantique, permettant au sujet d'affecter la réalité. C'est le résultat de l'interaction qui est «mesuré», et non la réalité indépendante du sujet. Une fonction d'onde (un état quantique) représente les degrés de croyance qu'un agent a sur les résultats possibles des «mesures».[1] Pour le QBiste, le concept primaire est celui d'expérience plutôt que de réalité, l'expérience étant comprise dans un sens subjectiviste comme une expérience personnelle plutôt que comme une partie de la réalité. En revanche, selon le réalisme contextuel, il n'y a pas de distance entre l'expérience primaire (non conceptualisée, immédiate) et la réalité. Le QBisme est une perspective en première personne, tandis que le réalisme contextuel est une position externaliste. Sur le plan épistémologique, nous pensons que le réalisme contextuel est compatible avec l'épistémologie de la «connaissance d'abord» (knowledge first epistemology) de T. Williamson [16]. Cette dernière est une épistémologie dans la perspective en troisième personne qui rejette l'épistémologie de l'«expérience d'abord» (experience first epistemology). Ainsi, le QBisme est une approche idéaliste (anti-réaliste) et subjectiviste plutôt que réaliste.[2] Ceci est confirmé par le fait que le QBisme s'inspire du pragmatisme et de l'empirisme radical de W. James, du pragmatisme normatif instrumental de D. Dewey et même du néo-pragmatisme postmoderne de R. Rorty, tout en ignorant pratiquement le pragmatisme normatif contextuel du second L. Wittgenstein. M. Bitbol et quelques autres auteurs ont proposé de considérer le QBisme dans les termes de la phénoménologie de E. Husserl et de la phénoménologie de Merleau-Ponty [15]. A notre avis, ces interprétations ne sont qu'approximatives et, à proprement parler, insatisfaisantes. En outre, la phénoménologie elle-même a besoin d'une correction réaliste. Le réalisme contextuel est une critique de la phénoménologie traditionnelle [19].

---

[1] Les probabilités sont plutôt interprétées au sens de Bruno de Finetti et Frank Ramsey. Pour le QBiste même les probabilités égales à l'unité sont des degrés de croyance.

[2] Selon nous, le «réalisme participatif» de C Fuchs (et M. Merleau-Ponty) n'est pas vraiment une position réaliste, mais plutôt celle corrélationniste, tandis que son «structuralisme normatif» [17] peut être réinterprété en termes de la notion de règle (norme) Wittgensteinienne. Notre position s'accorde avec celle de D. Glick. Le philosophe américain a proposé un «réalisme normatif perspectif» comme une position alternative au réalisme scientifique standard qui permet de comprendre le QBisme de manière réaliste. À la différence du réalisme scientifique standard, le réalisme normatif perspectif prend en compte la dimension perspective et celle normative [18].

### 4. Un principe de quantification de l'information

Nous analysons le «principe conceptuel fondamental de la mécanique quantique», proposé par le physicien autrichien A. Zeilinger [2] et montrons qu'il s'agit d'un principe idéaliste, qui devrait être remplacé par le principe réaliste de la contextualité.[3]

Le scientifique propose le principe suivant: (1) «An elementary system carries 1 bit of information» (un système élémentaire porte1 bit d'information) [2, p. 635]. Il note toutefois que cette formulation peut être considérée comme une définition du système physique minimal. De manière équivalente – et c'est important pour comprendre notre position – le principe fondamental peut être reformulé en termes linguistiques: (2) «An elementary system represents the truth value of one proposition» (le système élémentaire représente la valeur de vérité d'une proposition) [2, p. 635]. Zeilinger appelle cette version du principe «un principe de quantification de l'information». Nous reformulerions (2) comme suit: (2') «Le système élémentaire est décrit/représenté par une proposition vraie». Le lien entre (1), (2) et (2') est évident: 1 bit d'information est exprimé comme une proposition vraie; par conséquent, le système auquel une proposition vraie est associée est celui qui «porte» 1 bit d'information.

Zeilinger déduit dudit principe les propriétés des systèmes quantiques. Nous acceptons cette déduction théorique. En même temps, nous pensons que la position philosophique du scientifique contient des contradictions et des confusions conceptuelles.

Par exemple, Zeilinger fait la distinction suivante entre les approches classique et quantique: «Therefore, while in a classical worldview, reality is a primary concept prior to and independent of observation with all its properties, in the emerging view of quantum mechanics the notions of reality and of information are on an equal footing» (Alors que dans la vision classique du monde, la réalité avec toutes ses propriétés est un concept primaire précédant l'observation et indépendant de celle-ci, dans la vision émergente de la mécanique quantique, les concepts de réalité et d'information sont sur un pied d'égalité) [2, p. 642].

La prémisse est fausse. Le concept de réalité n'est pas ambigu, il est le même pour tout domaine scientifique: «reality is just what it is» (la réalité est simplement ce qu'elle est) [5, p. 22]. Par définition, elle ne peut pas être «sur un pied d'égalité» avec l'information qui, si elle est comprise comme connaissance, appartient à la catégorie de l'idéal (l'information est connaissance de la réalité et non une partie (particulière) de la réalité).[4] Zeilinger se contredisant lui-même écrit: «We have knowledge, i.e., information, of an object only through observation» (Nous n'avons de connaissance, c'est-à-dire d'information, sur un objet que par l'observation) [2, p. 633].

«La vision classique du monde» est tout à fait correcte en considérant la «réalité» comme un concept primaire. Une autre chose est que le réalisme métaphysique traditionnel, qui affirme le caractère «(pré)donné» de la réalité externe avec des propriétés déterminées – des propriétés internes des objets «externes» autonomes (décontextualisés, absolutisés) – est une position fausse.

---

[3] Déjà après avoir écrit cet article, nous avons appris que la critique du principe d'information de Zeilinger et Bruckner a été proposée indépendamment par T. Bilban, cependant, sur la base de la position phénoménologique de Husserl [20-21]. Voir également une analyse critique de l'usage du concept d'information dans l'interprétation informationnelle de la mécanique quantique, proposée par C. Timpson [22-23].

[4] Cet article traite de la notion épistémique d'information en tant que connaissance. La notion d'information, qui apparaît dans la théorie de l'information de C. Shannon, a un sens technique [22-23]. Dans un sens technique, le terme est utilisé par C. Rovelli dans son interprétation relationnelle de la mécanique quantique et sa dérivation des postulats informationnels [24].

On ne peut parler de ce type de réalité censée et significative en elle-même que de manière secondaire – comme d'un domaine de réalité déjà connue, «apprivoisée» (conceptualisée). Comme si les choses qui sont déjà connues (identifiées) portaient des informations sur elles-mêmes.

En fait, on peut être d'accord avec Zeilinger lorsqu'il écrit que «in physics we cannot talk about reality independent of what can be said about reality» (en physique, on ne peut pas parler de la réalité indépendamment de ce que l'on peut dire de la réalité), mais seulement si cette formulation est comprise non pas dans le sens de l'idéalisme linguistique (le langage est primaire, la réalité est secondaire et correspond au langage ou est construite par lui) ou du corrélationnisme (le langage en lui-même et la réalité en elle-même n'ont pas de sens ou sont en tout cas inconnaissables; il n'y a que des «correlations» indépassables du langage et de la réalité; nous n'avons accès qu'aux choses exprimées dans le langage, choses-pour-nous, pas aux choses elles-mêmes), mais comme une tautologie: nous ne pouvons vraiment pas parler (ou penser) de la réalité (des choses réelles) si nous ne pouvons pas en parler (penser), c'est-à-dire si nous n'appliquons pas les moyens appropriés pour l'identifier (les identifier). Comme le dit A. Peres, «unperformed experiments have no results» (les expériences qui n'ont pas été réalisées n'ont aucun résultat) [25]. C'est la vision logique (thérapeutique) du réalisme contextuel. Le langage n'est en «correlation» avec la réalité que dans la mesure où il est capable de la décrire de manière adéquate dans son contexte, de la comprendre. C'est de cette manière – en tant que critique du réalisme métaphysique et du corrélationnisme, et non au sens néo-kantien – que nous proposons également d'interpréter les paroles de Bohr citées par Zeilinger: «It is wrong to think that the task of physics is to find out how Nature is. Physics concerns what we can say about Nature» (il est faux de penser que la tâche de la physique est de découvrir ce qu'est la Nature. La physique s'occupe de ce que nous pouvons dire sur la nature) (cité dans [2, p. 637]). En fait, ce qu'est la nature (la réalité, les choses mêmes) est découvert dans nos jugements contextuellement corrects sur la nature. Notre langage ordinaire ainsi que notre langage scientifique est fait pour ça.

### 5. Du principe informationnel au principe contextuel

La déduction informationnelle de Zeilinger suppose l'interpénétration de l'information et de la réalité (il y a ici, à notre avis, un parallèle avec l'idéalisme objectif de Hegel) et la conservation de l'information. Par conséquent, l'absence d'information équivaut à l'absence de réalité: d'où la quantification des quantités physiques, le caractère objectif des probabilités quantiques, l'impossibilité d'introduire des paramètres cachés, la «non-localité» de la corrélation quantique, le principe de complementarité, etc. Le problème est que les systèmes quantiques, portant en eux-mêmes des informations sur eux-mêmes, s'avèrent des systèmes autonomes (absolutisés).

Nous proposons plutôt de considérer les systèmes quantiques et leurs caractéristiques comme identifiables dans leur contexte. L'ontologie quantique (en fait toute ontologie) et la connaissance sont contextuelles [12] (voir aussi [26]).[5] L'identification (cognition) dans le contexte est exprimée en tant que connaissance propositionnelle. L'expression élémentaire de la connaissance est la proposition. Cela signifie qu'il découle du principe de contextualité qu'un système physique élémentaire peut être identifié sous la forme d'une seule proposition (1 bit d'information). Pour Zeilinger, au contraire, un système physique représente un contenu propositionnel et (porte en lui) une information. En même temps, se contredisant lui-même, il explique que des énoncés tels que «un système "représente" la valeur de vérité d'une proposition ou "porte" un bit d'information» doivent être compris comme des énoncés «sur ce qui peut être dit sur les résultats possibles d'une mesure» [2, p. 635].

Les propriétés spécifiques des systèmes quantiques sont une conséquence de l'impossibilité de parler d'eux sans référence aux moyens de leur identification et, donc, au contexte dans lequel ces moyens sont appliqués (en général, selon le théorème de P. Destouches-Février, les théories, au sein desquelles les phénomènes ne sont pas séparables de la manière dont on y accède, sont essentiellement probabilistes [30]). En particulier, l'hypothèse de non-localité n'est pas nécessaire pour expliquer la corrélation quantique.[6] Les événements quantiques corrélés sont liés les uns aux autres de manière causale. Mais il ne s'agit pas de causalité classique, mais de causalité quantique, exprimée par une fonction d'onde intriquée. Telle ou telle corrélation particulière n'apparaît pas dans la mesure ; dans la mesure, elle est identifiée dans le contexte [11, 12].[7] Contrairement au principe de quantification de l'information proposé par Zeilinger, le principe de contextualité l'explique de manière réaliste.

---

[5] La nature essentiellement contextuelle de la physique quantique est soulignée par M. Bitbol [15, 27]. Récemment, F. Grangier a également défendu l'idée que «les propriétés/modalités appartiennent au système dans son context» [28]. Sur la contextualité dans l'interprétation de Copenhague de la mécanique quantique voir [29]. Quant à nous, nous parlons de contextualité au sens du second Wittgenstein et du réalisme contextuel de Benoist [9, p. xii–xiii] (voir la citation ci-dessus au paragraphe 2).

## 6. Conclusion

Ainsi, selon nous, la position philosophique de Zeilinger peut et doit être inversée, remise sur ses pieds. En conséquence, il s'avère que le principe fondamental de la physique quantique n'est pas le principe d'information, mais de contextualité. Ce n'est que dans une perspective contextuelle que la déduction informationnelle des propriétés et des phénomènes quantiques proposée par Zeilinger prend un sens véritablement réaliste. N. D. Mermin a écrit: «New interpretations appear every year. None ever disappear» (de nouvelles interprétations apparaissent chaque année. Aucune ne disparaît) [31, p. 8]. Notre vision de la mécanique quantique – le réalisme quantique contextuel (RCQ) – n'est pas une autre interprétation, mais révèle la signification réelle de la mécanique quantique, rendant ses interprétations inutiles [32].

---

[6] C'est encore un point commun entre notre réalisme quantique contextuel et le QBisme: la mécanique quantique peut etre considérée comme une théorie locale.

[7] Déjà après avoir écrit cet article, nous avons appris que Grangier défend un point de vue proche, mais pas le même que le nôtre, sur les corrélations quantiques [28].

# On a foundational conceptual principle of quantum mechanics

## F. Laurent & F.-I. Pris

We argue that Anton Zeilinger's «foundational conceptual principle» for quantum mechanics according to which an elementary system carries one bit of information is an idealistic principle, which should be replaced by a realistic principle of contextuality. Specific properties of quantum systems are a consequence of impossibility to speak about them without reference to the tools of their observation/identification and, consequently, context in which these tools are applied. In particular, the assumption of non-locality is not required to explain quantum correlation. Correlated quantum events are related with each other in a causal way. This is not classical, but quantum causality expressed by an entangled wave function. This or that particular correlation does not arise in measurement; in measurement it is identified in context. In contrast to Zeilinger's informational principle, the principle of contextuality explains it realistically.

*Keywords:* quantum mechanics, information, idealism, realism, context, rule-following problem, quantum correlation

Frédérique Laurent
(traductrice littéraire, France)

François-Igor Pris
(docteur en philosophie, docteur en physique théorique, chercheur principal à l'Institut de philosophie de l'Académie nationale des sciences du Belarus)

Leading researcher at the Institute of philosophy of the National academy of sciences of the Republic of Belarus (Minsk, Belarus), PhD in philosophy, PhD in theoretical physics.
frigpr@gmail.com

+ 375 44 5147155